\documentclass[aps,reprint]{revtex4-2}
\usepackage{graphicx}
\usepackage{dcolumn}
\usepackage{bm}

\usepackage[utf8]{inputenc}
\usepackage[T1]{fontenc}
\usepackage{mathptmx}
\usepackage{etoolbox}

\usepackage{color}
\usepackage{amsmath}
\usepackage{float}

\makeatletter
\def\@email#1#2{%
 \endgroup
 \patchcmd{\titleblock@produce}
  {\frontmatter@RRAPformat}
  {\frontmatter@RRAPformat{\produce@RRAP{*#1\href{mailto:#2}{#2}}}\frontmatter@RRAPformat}
  {}{}
}%
\makeatother
\begin{document}

\preprint{AIP/123-QED}
\title{Deterministic positioning accuracy of B centers in hBN by electron irradiation}
\author{A. N\'u\~nez Marcos, C. Arnold, J. Barjon, S. Buil, J.-P. Hermier, A. Delteil}

 \affiliation{Universit\'e Paris-Saclay, UVSQ, CNRS, GEMaC, 78000, Versailles, France}

\date{\today}

\begin{abstract}

Hexagonal boron nitride (hBN) is a promising material platform for quantum technologies, enabling both the realization of nanophotonic devices and the hosting of point defects that act as high-quality single-photon emitters (SPEs). Among these defects, B centers, which reproducibly emit at 436 nm, can be generated individually at predefined positions using electron beam irradiation, a crucial capability for top-down quantum photonics. Yet, the spatial accuracy of emitter creation strongly conditions the potential applications. Here, we generate arrays of individual B centers in a scanning electron microscope (SEM) with in-situ cathodoluminescence (CL) monitoring. We characterize the sources of random deviations from the target positions to identify the limitations to their deterministic positioning. In particular, we measure the distance between the electron beam and the emitter at the moment the emitter is created based on CL mapping. This approach provides a direct estimation of the spatial extent of the beam-emitter interaction. We find this contribution to be the main constraint to the emitter fabrication accuracy, yielding a standard deviation of 27~nm. Additional contributions, primarily from the finite density of UV-emitting carbon dimer precursors, increase the overall positional standard deviation to 40~nm. Our results highlight the possible improvements to the B center creation technique to reach sub-10~nm accuracy.
\end{abstract}

\maketitle

Optically active deep defects in wide-bandgap semiconductors are widely regarded as key building blocks for quantum technologies, offering robust single-photon emission and opportunities for top-down integration in photonic structures~\cite{Aharonovich2016, Zhang20}. Among these materials, hexagonal boron nitride (hBN) has emerged as a van der Waals host for a diverse range of color centers with exceptional optical properties, including high brightness and stability at room temperature~\cite{Tran16, Martinez16, Vogl25}. Notably, one family of emitters -- the B centers -- can be locally generated using an electron beam~\cite{Shevitski19, Fournier21, Gale22}, enabling reproducible single-photon emission at 436~nm with excellent quantum optical figures of merit at both ambient and low temperature \cite{Horder22, Fournier23PRA, Gerard24, Gerard26}. 

Importantly, hBN as host material offers the possibility to realize monolithic photonic structures in which the SPEs can be a posteriori integrated in a top-down approach \cite{Gerard23,Spencer23}. However, performance and scalability require a high control on the spatial positioning of the emitter. This is particularly crucial for resonant structures such as microcavities and plasmonic resonators, where deviations from the optimal position rapidly degrades the performances. 

Here, we investigate the positioning accuracy in the fabrication of individual B centers by electron irradiation in a SEM. The local activation of B centers involves the modification of native UV-emitting color centers (carbon dimers $\mathrm{C_B C_N}$~\cite{Mackoit19, Li22, Plo25}) at the same position through a one-to-one conversion process  \cite{Nunez26}. This conversion can be monitored in real time using cathodoluminescence (CL) \cite{Shevitski19, Roux22, Nedic24, Hou25}, allowing to stop the irradiation upon appearance of the B center optical signature. In this work, we exploit this ability to generate arrays of individual emitters. We then use photoluminescence (PL) to characterize the emitter arrays and investigate the deviations of the B center positions with respect to the target locations. We then identify the various contributions to the nonideality of the emitter arrays, which we separately investigate using a combination of CL mapping experiments and simulations. Our study focuses on the in-plane generation accuracy. In turn, the vertical distribution of the generated emitters is bounded by the crystal interfaces and can be as low as 3~nm~\cite{Liu25}.

 \begin{figure}[!h]
 \includegraphics[width=\linewidth]{ 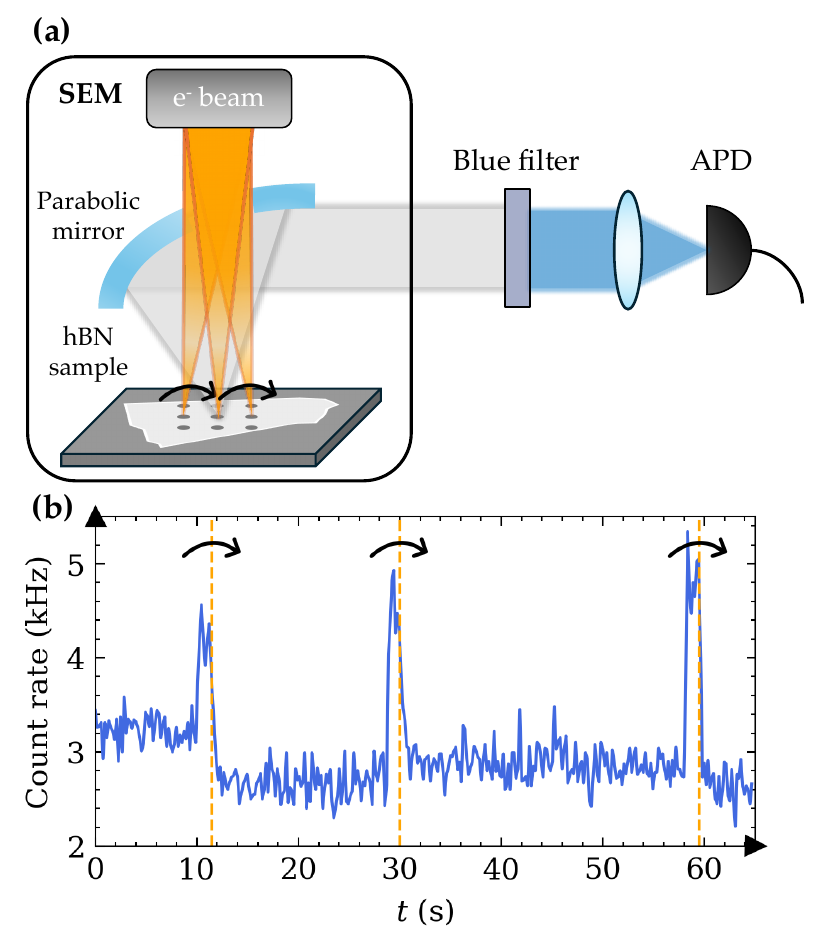}
 \caption{\label{Experimental setup} (a) Scheme of the electron irradiation setup with in-situ monitoring of single B center creation. (b) Typical time trace recorded during the irradiation, with upward steps corresponding to the creation of a SPE, immediately followed by the e-beam displacement to the next site, as indicated by the vertical lines.}%
 \end{figure}

\begin{figure*}[]
 \centering
 \includegraphics[width=\linewidth]{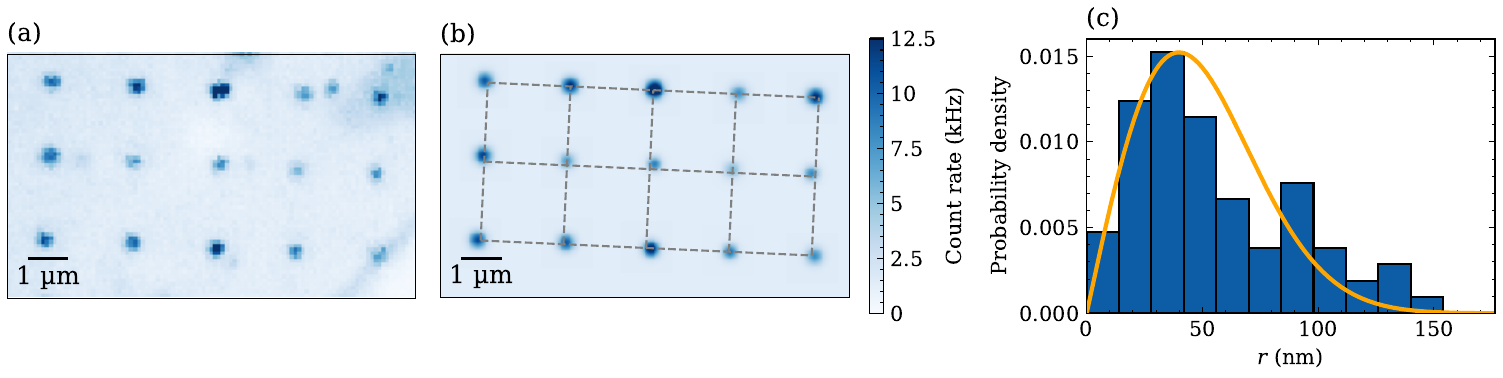}%
 \caption{\label{PL reconstruction} (a) PL confocal map of a $3 \times 5$  B-center array measured with a $200$~$\mu$W pulsed laser with a piezoelectric step of $100$~nm on a $86$~nm-thick flake at room temperature. (b) Blue shading: 2D Gaussian fit of the PL data in (a). Gray dashed lines: target grid with even spacing of (c) Histogram of the deviation between the target and real positions of the B centers. Orange curve: Rayleigh distribution fit with its maximum value at $39.9 \pm 2.8$~nm.} %
 \end{figure*}

The sample consists of a single crystal of high-pressure, high-temperature grown hBN \cite{Taniguchi07} exfoliated on a SiO$_{2}$(300~nm)/Si substrate. After acetone and isopropanol cleaning, the sample is annealed at $1000 \ ^{\circ}$C under 200~sccm nitrogen (N$_{2}$) flow during 1~h and left inside the oven during 3~h for cooling. This technique enhances the formation of UV-emitting color centers. The B centers are created by $3$~kV irradiation and a current of $I = 0.023$~nA in high vacuum conditions ($P < 10^{-4}$ Pa) at room temperature. CL emission is filtered with a blue bandpass filter with a passband ranging from 420 to 460~nm and detected by an avalanche photodiode (APD).

Fig.~\ref{Experimental setup}a depicts the experimental setup that enables real-time CL monitoring of individual B-center generation. The creation of a B-center array is performed by successively placing the beam to each setpoint of a predetermined square array and monitoring the CL count rate until it signals the activation of a B center. As soon as a characteristic upward step is observed, the electron beam is sent to the next setpoint using the built-in electrical control of the deflection coils (Fig.~\ref{Experimental setup}b).

After electron irradiation, the sample is characterized in photoluminescence (PL) in a scanning confocal microscope, where a pulsed laser emitting at $405$~nm with circular polarization is used to excite the emitters. An example of PL confocal scan of a $3 \times 5$ array with $2~\mu$m nominal step is shown on Fig.~\ref{PL reconstruction}a. The second-order correlation function of most of the spots presents single-photon emission with $g^{(2)}(0)<0.5$~\cite{Nunez26}. Due to diffraction, the color centers appear as Gaussian spots of standard deviation $148 \pm 34$~nm. Note that the localization precision of the emitter can be substantially smaller than this spot size depending on the signal over noise ratio and the data analysis method used to estimate the emitter position~\cite{Liu17, Sartison17, Laurent2026, LamraniAlaoui2026}. To estimate the position of the spot center, we fit each one with a 2D Gaussian function, which provides their location with an uncertainty $\sigma_\mathrm{PL} = 8$~nm. This value is lower than the spot width, and is only limited by the signal over noise ratio and the positioner resolution (in our case, close-loop positioners with 4~nm accuracy). The localization accuracy from a Gaussian fit to the PL spot can reach below 1~nm under favorable photon-count conditions.~\cite{Liu17, Sartison17}. The fit result associated with the confocal scan of Fig.~\ref{PL reconstruction}a is shown on Fig.~\ref{PL reconstruction}b. Afterwards, following Chen \textit{et al.}~\cite{Chen19}, we infer the target positions from the map by fitting a rectangular grid to the emitter positions. The free parameters involved are the pitch in the $x$ and $y$ directions, the spatial offset, the grid shear and the in-plane rotation angle. This method eliminates the need for absolute calibration and natively accounts for all potential systematic errors of both the SEM and PL imaging systems, thereby allowing for a precise analysis of the purely stochastic contributions to the SPE fabrication process.

We repeat this process on 6 similar grids, which encompass a total of 75 SPEs. Fig.~\ref{PL reconstruction}c shows the histogram of the distance between the target position and the actual location of each emitter (blue bars). The orange curve is a fit to the distribution by a Rayleigh function of the form
 \[
 R(r) = \dfrac{r}{\sigma^2} \exp \left(
 		-\dfrac{r^2}{2\sigma^2}
 		\right)
 \]
which is the radial probability distribution associated with a 2D Gaussian distribution
 \[
G(r) = \dfrac{1}{2 \pi\sigma^2} \exp \left(
 		-\dfrac{r^2}{2\sigma^2}
 		\right).
 \]
While the probability density $G(r)$ is maximal at the center, the radial probability density $R(r)$ is maximal at $r=\sigma$. This difference arises from the increasing area available at increasing radius: an annulus of radius $r$ and width $dr$ has an area $2\pi r\,dr$, such that $R(r)=2\pi r G(r)$.

The fit yields $\sigma_\mathrm{tot} = 39.9 \pm 2.8$~nm, which is the standard deviation of the Gaussian distribution along either spatial coordinate and, equivalently, the most probable radial distance from the center. For a 2D Gaussian, the probability of finding a single emitter within a radius $2\sigma$ is $\simeq 86 \%$. The obtained value of $\sigma_\mathrm{tot}$ is comparable to the best reported results of spatially controlled generation of color centers in diamond~\cite{Chen17, Chen19, Kim25, Cheng25}, silicon carbide~\cite{Wang17, He23} and silicon~\cite{Hollenbach22}, and rare-earth ions in YAG crystals~\cite{Groot19}. To gain insight on the possible improvements, we now turn to the analysis of the contributions to the finite spread of the relative positions.

The total average deviation $\sigma_\mathrm{tot}$ estimated from the histogram of Fig.~\ref{PL reconstruction}c can be described as originating from four main independent contributions, which we independently characterize in the following: the accuracy $\sigma_\mathrm{target}$ with which the electron beam control system can be positioned on the target site, the precision with which the emitter position is estimated from the PL map $\sigma_\mathrm{PL}$, the width of the spatial distribution of the electron beam energy deposition in the crystal $\sigma_{e^{-}}$ and an additional contribution originating from the finite density of the UV center precursors in the irradiated crystal $\sigma_{n}$. The latter contribution is specific to our in-situ monitored generation protocol. Its calculation and its treatment as an independent contribution will be discussed later in the text, as well as in the supplementary information. The total spread of the emitter relative position then writes

\begin{equation}
\label{sigma_tot_eq}
\sigma_\mathrm{tot}^2 = \sigma_\mathrm{target}^2 + \sigma_\mathrm{PL}^2 + \sigma_\mathrm{e^-}^2 + \sigma_n^2
\end{equation}

Among the terms contributing to $\sigma_\mathrm{tot}$ in Eq.~\ref{sigma_tot_eq}, two can be easily estimated: $\sigma_\mathrm{PL}$, which we infer to be $8 \pm 4$~nm from the fits of the PL maps, and $\sigma_\mathrm{target}$, the accuracy of the electron beam positioning system. Based on SEM image of irradiated arrays (see Supplementary Information section~S2), we find $\sigma_\mathrm{target} = 18 \pm 1$~nm.

\begin{figure}[!t]
 \includegraphics[width=\linewidth]{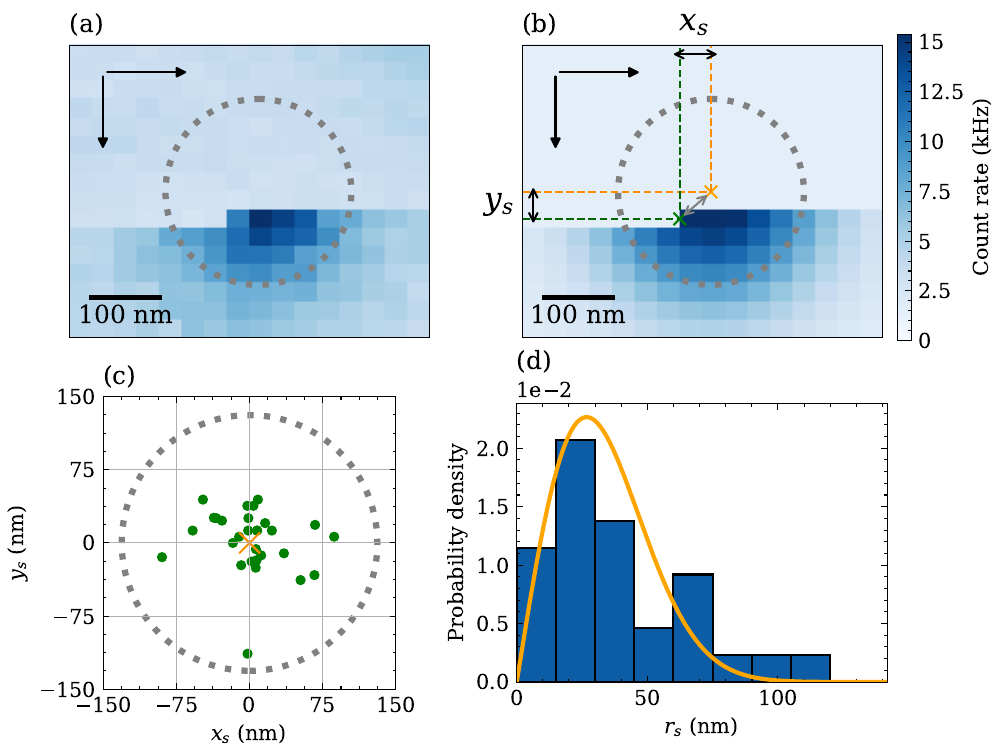}%
 \caption{\label{B center generation fit} (a) CL map of a B center activated during the scan. The arrows indicate the electron scan direction and the gray dotted circle depicts the FWHM of the CL spot, estimated to be $260 \pm 15$~nm (see Supplementary Material section S4). (b) Fit of the CL map in (a) using Eq.~\ref{fit_CL}. Orange lines: center position of the CL spot. Green lines: position of the electron beam at the moment of the B center activation. (c) Spatial distribution of the electron beam relative positions during the activation of 29 B centers. (d) Distribution of the distance between the B centers and the electron beam position at activation. Orange line: Rayleigh distribution fit with a maximum value at $26.8 \pm 2.5$~nm.}%
 \end{figure}

In turn, $\sigma_{e^-}$ describes the distance between an emitter and the position of the electron beam that triggers its activation. The larger $\sigma_{e^-}$, the broader the area in which the electron beam can activate an emitter at a given dose. To characterize the distribution of the emitter-beam distance, and therefore $\sigma_{e^-}$, we perform an additional experiment in CL mapping mode. In this regime, we continuously scan a given area of the sample while recording the CL signal at each position. If a B center is present on the scanned area, it appears as a Gaussian spot on the CL map, with a certain spread characterized by a standard deviation $\sigma_\mathrm{CL} = 111 \pm 7$~nm (see Supplementary Information section~S4). We emphasize that $\sigma_\mathrm{CL}$ describes the distance at which the beam can excite the emitter, not the distance at which it can activate it. Occasionally, a B center is activated during the scan. It then appears as a truncated Gaussian -- a characteristic shape that has been reported in prior work~\cite{Shevitski19, Hou25, Nunez26}. An example of such activation event during scanning is shown on Fig.~\ref{B center generation fit}a. Remarkably, this observation makes it possible to correlate the relative position of the electron beam and the emitter at the moment the B center is activated: the emitter position is given by the center of the Gaussian, while the pixel at which the signal appears marks the position of the electron beam at activation. This process is independent of all the other sources of inaccuracy included in Eq.~\ref{sigma_tot_eq}.

We perform an estimation of this relative position by fitting the CL map to a truncated 2D Gaussian of the form:

\begin{equation}
\label{fit_CL}
f(x, y) = A\ \exp \left(
\frac{(x - x_{0})^{2} +(y - y_{0})^{2} }{2 \sigma_\mathrm{CL}^{2} }
\right)
M(x, y) + D 
\end{equation}
where $A$ is the signal maximum intensity, $(x_{0}, y_{0})$ is the center position of the CL spot, $\sigma_\mathrm{CL} $ is the standard deviation of the Gaussian, $D$ is the background signal, and $M(x,y)$ is a binary mask describing the activation, which reads:

\begin{align*}
M(x, y) &= \begin{cases}
	1 &\text{if } y > y_{0} + y_{s} + \delta y/2 \\
	  &\text{or if } y > y_{0} + y_{s} - \delta y/2 \text{ and } x > x_{0} + x_{s} \\
	0 &\text{elsewhere}
\end{cases}
\end{align*}
where $x_{s}$ ($y_{s}$) is the horizontal (vertical) offset between the electron beam at activation and the CL spot center, and $\delta{y}$ is the vertical pixel size of the scan.

As a representative example, the fit results of the truncated CL spot of Fig.~\ref{B center generation fit}a is shown on Fig.~\ref{B center generation fit}b. To perform a statistical analysis, we repeat this process on 29 CL spots. Fig.~\ref{B center generation fit}c shows the spatial distribution of the activation position with respect to the center of the CL spots. The radial distance of these points are plotted as a histogram on Fig.~\ref{B center generation fit}c, and are fitted to a Rayleigh distribution, which provides $\sigma_{e^-} = 26.8 \pm 2.5$~nm as the most probable distance at which the B centers are activated -- thereby yielding a direct estimation of the beam-emitter interaction area.

This approach allows for a direct estimation of the spatial extent of the emitter-beam interaction. Here, the measured distribution is narrower than the size of the CL spot, which is Gaussian with standard deviation $\approx 110$~nm (see Supplementary Material section S4). In hBN, this value is limited by the diffusion of the excitons inside the crystal (gray dotted line on Fig.~\ref{B center generation fit}). In turn, the value of $\sigma_{e^-}$ is of the same order as the standard deviation of the calculated beam broadening during the electron deceleration inside hBN calculated using CASINO Monte-Carlo simulations~\cite{CASINO}, which provide $14.7 \pm 2.1$~nm as we expose in the Supplementary Material section~S1. This observation is consistent with a dose-dependent activation mechanism triggered by primary electrons~\cite{Roux22}. The discrepancy can be explained by the non-ideal beam focusing in our experiment and the questionable accuracy of Monte-Carlo simulations at low voltage. Nevertheless, the simulations indicate that reducing the crystal thickness below 35~nm would only minimally broaden the beam, reducing this contribution to less than 8~nm for a crystal thickness of 25~nm.

The remaining term $\sigma_n$ can be estimated from an effective model. As we show in detail in the Supplementary Information section~S3, this term is associated with the finite surface density $n_s$ of UV-emitting precursors in the crystal, and can be expressed as $\sigma_n = 1/\sqrt{2 \pi n_s}$ to an excellent approximation. It is inherently associated with the in-situ monitoring process: if the precursor density is high, the electron beam will rapidly convert one of the UV centers overlapping with the interaction volume. In contrast, if the density is very low, in most cases no UV center lies directly near the beam center, and the irradiation will continue until an emitter further in the tails of the energy deposition distribution can be activated, artificially increasing the statistical distance between the electron beam position and the emitter location. The in-situ monitoring indeed guarantees that the irradiation will not stop until a UV center has been converted to a B center. Following the method described in ref.~\cite{Roux22}, we have estimated independently the number of precursors per unit area to be $n_s = 1.95 \pm 0.32 \times 10^{11}$~cm$^{-2}$, corresponding to a volume density of $n_v = 3.6 \pm 0.6 \times 10^{15}$~cm$^{-3}$, a value close to the result of ref.~\cite{Roux22} in a similar sample. This estimation is detailed in the Supplementary Material section~S5. This density corresponds to an additional contribution of $\sigma_n = 22.6 \pm 3.7$~nm to the activation accuracy. A decrease of $\sigma_n$ can be obtained through a higher UV center density: a value of $n_v = 2 \times 10^{16}$~cm$^{-3}$ would reduce $\sigma_n$ below 10~nm.

Based on the separate estimation of the four terms, we can estimate the expected accuracy based on Eq.~\ref{sigma_tot_eq}. We find $\sqrt{\sigma_\mathrm{target}^2 + \sigma_\mathrm{PL}^2 + \sigma_\mathrm{e^-}^2 + \sigma_n^2} = 40.2$~nm, in excellent agreement with the value of $\sigma_\mathrm{tot} = 39.9 \pm 2.8$~nm obtained from the PL maps.

In summary, we demonstrated the generation of B center arrays with an accuracy of $40$~nm in the in-plane directions. This precision is already enough for a wide variety of nanophotonic applications, opening the way to fully deterministic top-down fabrication of photonic chips in hBN. Our analysis allowed us to identify the factors responsible for the deviations from the ideal target position, with the main contributions being the spatial distribution of the energy deposition by the electron beam in the crystal and the finite density of the B center precursors. The former could be improved by modified electron irradiation conditions and a thinner crystal. The latter would require a higher concentration of carbon dimers, which can be obtained by intentional carbon doping of hBN crystals during growth~\cite{Pelini19, Williams25, Alemoush26}. A high concentration of carbon dimers in a thinner hBN crystal combined with improved electron beam parameters would allow for a generation accuracy below $10$~nm.

\section*{Acknowledgments}
This work is supported by the French Agence Nationale de la Recherche (ANR) under reference ANR-25-CE47-3104. This work received government funding under France 2030 (QuanTEdu-France) with reference number ANR-22-CMAS-0001. The Authors acknowledge K. Watanabe and T. Taniguchi for providing hBN crystals.

\pagebreak
~
\newpage

\onecolumngrid
\begin{center}
  \textbf{\large Supporting Information\\~\\ Deterministic positioning accuracy of B centers in hBN by electron irradiation}\\[.2cm]
  A. N\'u\~nez Marcos, C. Arnold, J. Barjon, S. Buil, J.-P. Hermier, A. Delteil\\[.1cm]
  {\itshape \small Universit\'e Paris-Saclay, UVSQ, CNRS,  GEMaC, 78000, Versailles, France. \\
{\color{white}--------------------} aymeric.delteil@usvq.fr{\color{white}--------------------} \\}

\end{center}

\setcounter{equation}{0}
\setcounter{figure}{0}
\setcounter{table}{0}
\setcounter{page}{1}
\renewcommand{\theequation}{S\arabic{equation}}
\renewcommand{\thefigure}{S\arabic{figure}}

\maketitle

\section*{S1. Simulation of the electron beam energy deposition}

\begin{figure}[h]
 \includegraphics[width=0.75\linewidth]{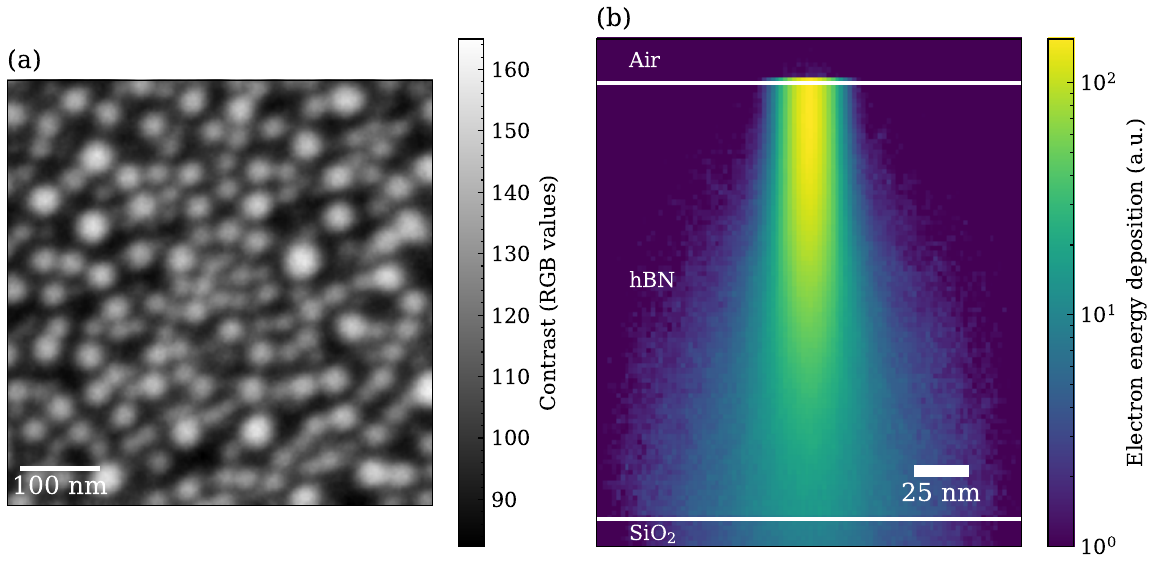}%
 \caption{\label{Electron irradiation simulations} \textbf{Electron beam simulations} (a) Example of SEM image of the standardized tin sample used to estimate the FWHM of the electron beam by FFT. (b) Vertical cross-section of the electron beam energy deposition from CASINO simulations with our B center irradiation conditions. Gaussian fit to the data yields an average standard deviation of $15 \pm 2$~nm.}%
 \end{figure}

Here, we evaluate the characteristic width of the energy deposition by the electron beam in the hBN crystal. We start by estimating the electron beam FWHM by Fast Fourier Transformation (FFT) image treatment \cite{SIRAF} on a standardized tin sample (S1937E). We obtain an electron beam FWHM value of $16$~nm, corresponding to a standard deviation of $\sigma = \mathrm{FWHM}/2.355 = 7.2$~nm. Note that this value is obtained under optimized focusing conditions and in the absence of the parabolic mirror, and is certainly underestimating the beam width used in the experiments in the main text. We perform CASINO simulations \cite{CASINO} of the electron trajectories and energy deposition in the sample given our irradiation conditions (acceleration voltage of $V = 3$~kV, electron current of $I = 23$~pA, beam FWHM of $16$~nm, hBN density of 2.18~g/cm$^{3}$ \cite{hBNdensity}, and a flake thickness of $d = 86$~nm). The results of the Monte-Carlo simulations are shown on Fig.~\ref{Electron irradiation simulations}b and c. The electron energy deposition has a cylindrical symmetry, with horizontal cuts being Gaussian to an excellent approximation for these irradiation conditions. The lateral extent of the energy deposition is estimated by performing 2D Gaussian fits of the calculated energy deposition at each horizontal plane, weighted by the total energy deposited at the corresponding $z$, and averaged. We obtain a standard deviation of $15 \pm 2$~nm. As discussed in the main text, this value is below the experimentally measured standard deviation of the distance between the electron beam position and the B center at the moment of activation ($\sigma_{e^-} = 26.8 \pm 2.5$~nm). Since the simulation only considers interaction between primary electrons and the crystal and considers ideal beam focusing, it provides a lower bound of the relevant interaction volume.

\section*{S2. Electron beam positioning accuracy}
 
\begin{figure}[h]
 \includegraphics[width=\linewidth]{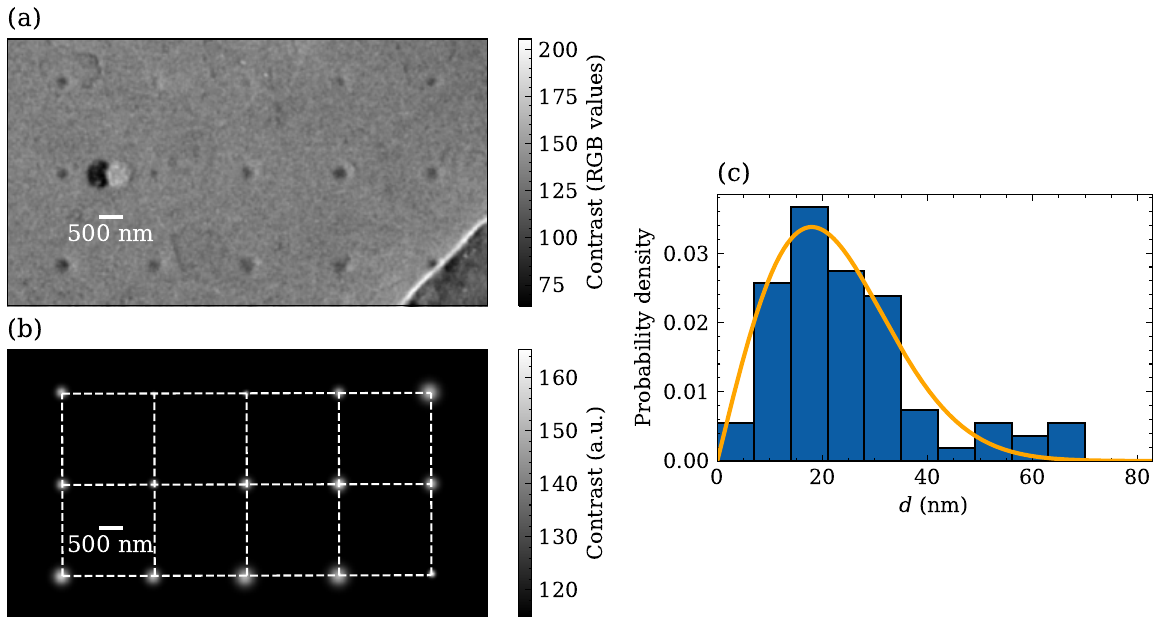}%
 \caption{\label{SEM reconstruction} \textbf{Electron beam positioning accuracy} (a) SEM image of the irradiated flake with the $3 \times 5$ grid presented on Fig.~2 of the main text. (b) White spots: fit results of the carbon contamination marks. White dashed line: target grid. (c) Blue bars: histogram of the deviation between the target positions and the centers of contamination marks. Orange curve: Rayleigh distribution fit yielding a maximum value at $18.0 \pm 0.9$ nm.}%
 \end{figure}

We explore the contribution of the scanning electron microscope to the global accuracy of the fabrication process. The electron beam position is defined by applying a certain voltage to the objective lenses of the microscope. Surface carbon contamination from the electron irradiation, seen as high contrast regions in the SEM image (Fig.~\ref{SEM reconstruction}a), reveals the beam position on the sample. By performing the same fitting process as for the PL grids (Fig.~\ref{SEM reconstruction}b), we can obtain the deviation of each spot from the ideal targeted grid (Fig.~\ref{SEM reconstruction}c). The average deviation value for the electron beam is found to be $\sigma_\mathrm{target} = 18.0 \pm 0.9$~nm.

\section*{S3. Contribution of the UV center density to the accuracy of the B center generation}

In this section, we present a simplified model that aims at evaluating the impact of the density of UV center precursors on the accuracy of the activation of a single B center by an electron beam. In our experimental protocol, the electron beam is kept switched on until the CL signal heralds activation of a blue center. Therefore, the conversion consists of the random selection of a UV center amongst the native UV centers, with a probability that depends not only on the electron beam spatial profile, but also on their surface density $n_s$.

\begin{figure}[h]
 \includegraphics[width=0.65\linewidth]{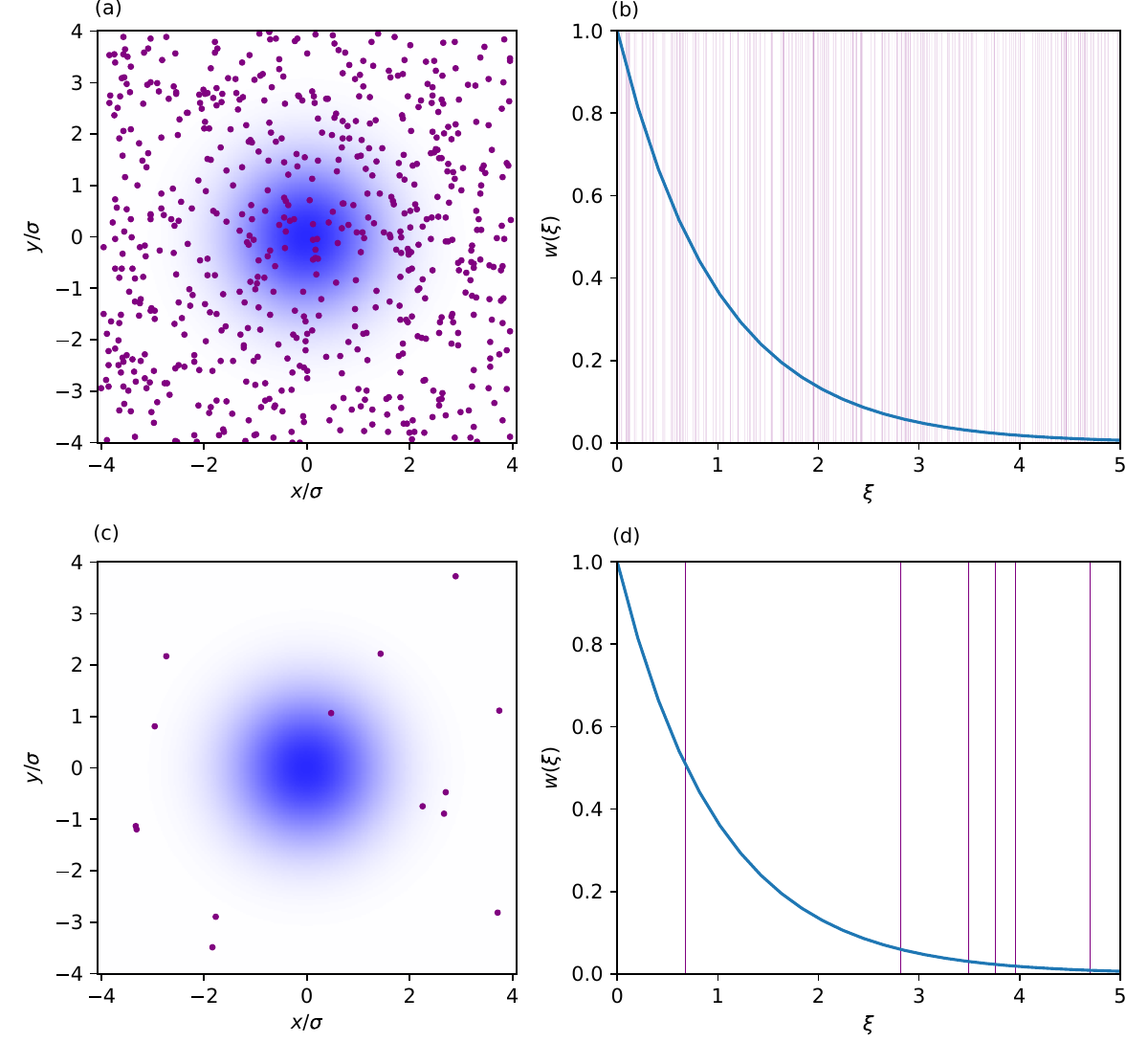}%
 \caption{\label{Simu1} (a) Example of numerically generated UV centers (violet dots) in the high density regime. The blue shading represents the 2D Gaussian selection weight. Here we took $n_\mathrm{UV}/\sigma_\mathrm{e^-}^2 = 10$. (b) Violet lines: same points in the transformed coordinates $\xi = r^2/2\sigma_\mathrm{e^-}^2$. The blue line represents the transformed selection weight function. (c) and (d): same as (a) and (b), but in the low density regime.}
 \end{figure}

We model the UV to B conversion by the random selection of one of the preexisting UV centers with a weight given by a 2D Gaussian probability density of standard deviation $\sigma_\mathrm{e^-}$, which models the interaction with the electron beam (blue shading on Fig.~\ref{Simu1}a and~c). The UV centers are assumed to have a surface density $n_s$ and uncorrelated random positions, thus realizing a 2D Poisson process (violet dots on Fig.~\ref{Simu1}a and~c).

To perform numerical simulations, we transform the problem to a 1D equivalent problem, where the random selection of a point in the 2D plane with weight $w(r) = e^{-r^2/(2 \sigma_\mathrm{e-}^2)}$ is equivalent to selecting a point on the half-line $\xi \geq 0$ with weight $w(\xi) = e^{-\xi}$, where we made the variable substitution $\xi = r^2/2\sigma_\mathrm{e-}^2$. This particular change of variable makes the UV center distribution a homogeneous 1D Poisson process on this half-line (Fig.~\ref{Simu1}b and~d).

\begin{figure}[h]
 \includegraphics[width=\linewidth]{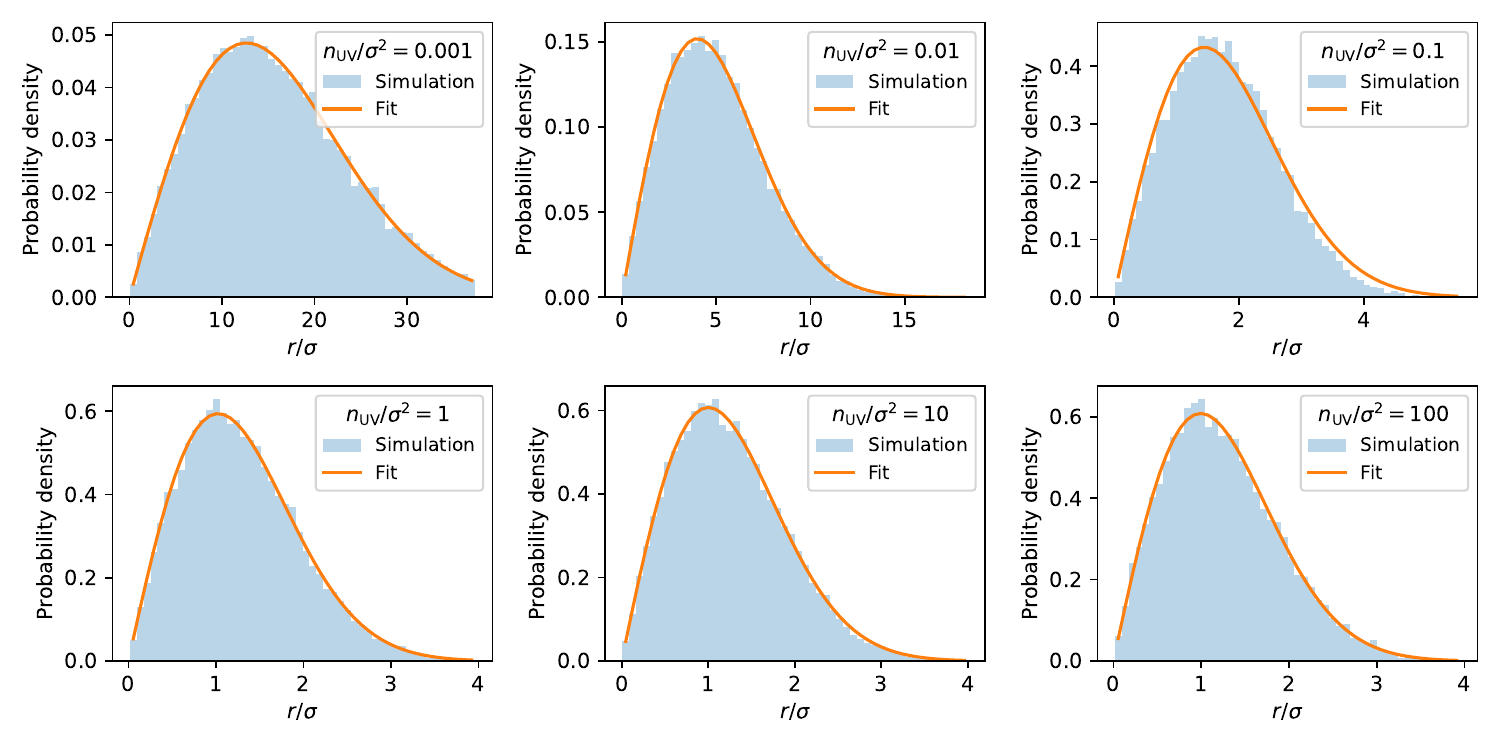}%
 \caption{\label{Simu2} Histogram of the distance between the selected UV center and the center of the 2D Gaussian selection weight function for different densities ranging from $n_s/\sigma_\mathrm{e^-}^2 = 10^{-3}$ to $n_s/\sigma_\mathrm{e^-}^2 = 10^2$.}
 \end{figure}

We simulate $10^4$ selection processes for a fixed value of $n_\mathrm{UV}$ to estimate the distribution of the selected UV center -- \textit{i.e.} converted to a blue center -- relatively to the electron beam center. The positions of the UV centers is shuffled every time. The process is repeated for several values of $n_\mathrm{UV}$ to investigate the effect of the UV center density. Fig.~\ref{Simu2} shows the results for different densities varying in a wide range, from the dilute limit ($n_s/\sigma_\mathrm{e^-}^2 = 10^{-3}$) to the dense limit ($n_s/\sigma_\mathrm{e^-}^2 = 10^{2}$). The results are converted back in the original spatial coordinate $r$. In all cases, the histogram can be well fitted to a Rayleigh distribution, testifying of a 2D Gaussian distribution of the activated center around the electron beam position. However, the histogram peaks at different distances depending on the density. While at high density, the maximum sits at $r/\sigma_\mathrm{e^-} \approx 1$, this most likely distance increases when the density decreases.

\begin{figure}[h]
 \includegraphics[width=0.6\linewidth]{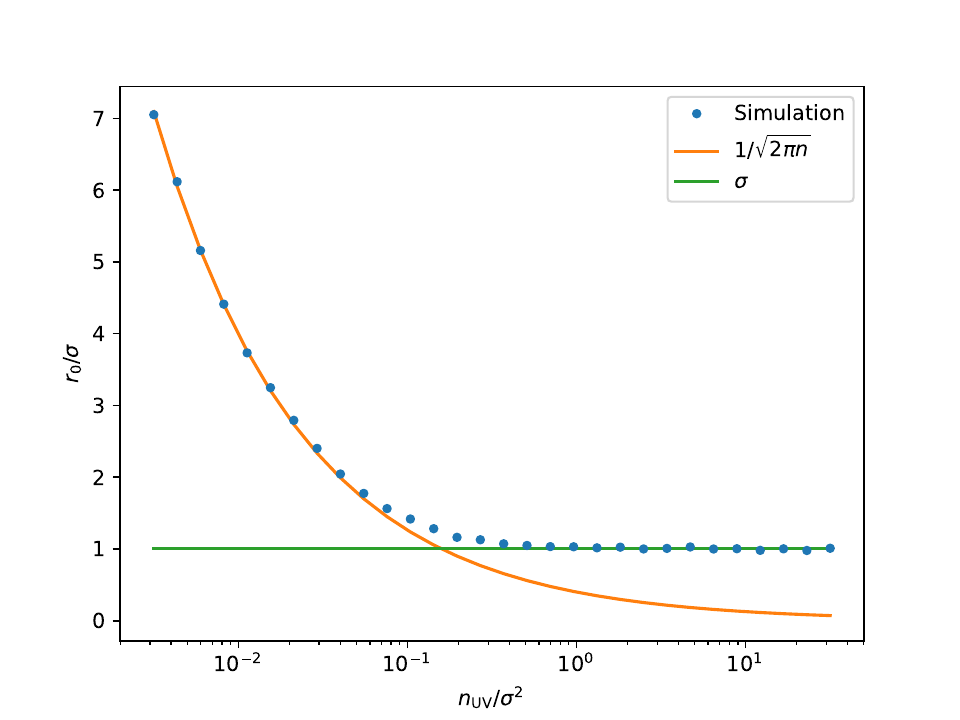}%
 
 \caption{\label{Simu3} Blue dots: simulated most likely distance $r_0$ as a function of the normalized density $n_s/\sigma^2$. The orange curve is calculated using the low-density limit $\sigma_n = 1/\sqrt{2 \pi n_s}$, and the green curve is the high-density limit value $r_0 = \sigma$. }
 \end{figure}

Fig.~\ref{Simu3} shows the most likely distance $r_0$ as a function of the normalized density $n_s/\sigma_\mathrm{e^-}^2$. Its asymptotic behavior in both dense and dilute limits can be captured by simple arguments. The selection process consists of associating a probability $P(i) = w(r_i)/\sum_j w(r_j)$ to each center $i$ and proceeding to a draw with respect to these probabilities.

In the dense limit, the numerator $\sum_j w(r_j) \approx n_\mathrm{UV} \int d^2 r\, w(r) = n_\mathrm{UV}$ becomes deterministic and the distribution of activated centers $P(i) = w(r_i)/n_\mathrm{UV}$ reflects the weight function. In other words, the converted UV centers realize an accurate sampling of the e-beam energy distribution. Their final distribution is therefore Gaussian, with a standard deviation $\sigma_\mathrm{e^-}$ (green curve on Fig.~\ref{Simu3}).

In the low limit in turn, there are in average no UV centers within the waist of the electron beam. Since the irradiation is not stopped until a conversion event occurs, the first conversion event is realized by the tail of the distribution. Due to the exponential decrease of the weight function, the UV center that sits the closest to the origin will be most likely selected. Therefore, the final distribution of the distance of the selected center matches the distribution of the closest neighbor in the 2D plane, which is a Gaussian distribution of standard deviation $\sigma_n = 1/\sqrt{2 \pi n_\mathrm{UV}}$ (orange curve on Fig.~\ref{Simu3}).

To summarize, the distribution of color centers activated by a random selection process having a 2D Gaussian distribution of width $\sigma_\mathrm{e^-}$ is itself Gaussian to an excellent approximation, with a density-dependent width. In the high-density limit, the distribution is given by that of the selection process, which is Gaussian in our particular experiment. In the low-density limit, the distribution is Gaussian because the conversion occurs by most likely selecting the UV center the closest to the electron beam (all the more so as the density decreases). In the general case, the distribution is well approximated by a Gaussian distribution of width $\sqrt{\sigma_\mathrm{e^-}^2 + \sigma_n^2}$, which captures both asymptotic behaviors. Conveniently, this makes it possible to treat the distance between the electron beam center position and the activated color center as a sum of two independent Gaussian random variables of respective standard deviations $\sigma_\mathrm{e^-}$ and $\sigma_n$. Note that, in the case of an irradiation with fixed dose (\textit{i.e.} without real-time feedback of the emitter creation), only the first contribution $\sigma_\mathrm{e^-}$ contributes, although at the price of a Poissonian distribution of the emitter number on each site.

\clearpage
\section*{S4. Line shape and size of the CL spots}
\begin{figure}[h]
 \includegraphics[width=0.75\linewidth]{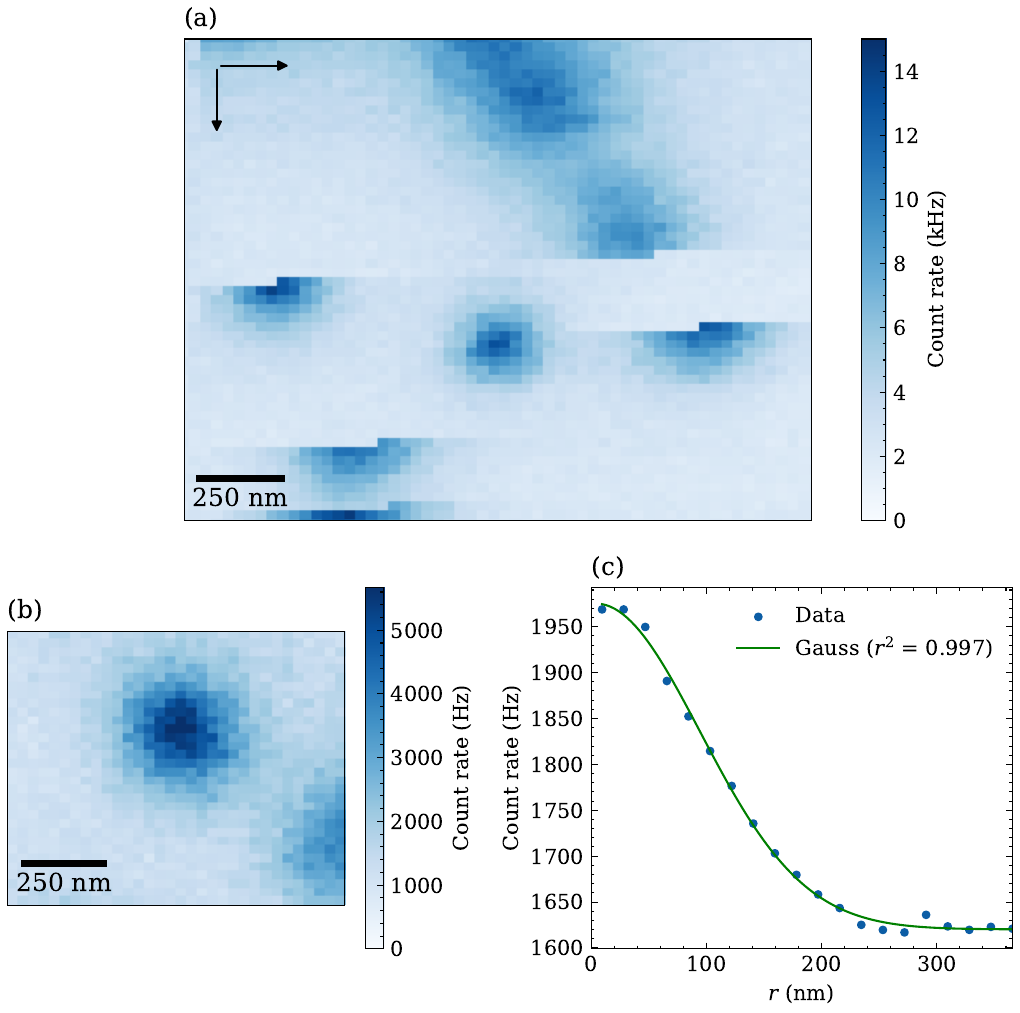}%
 \caption{\label{CL shape} (a) CL map with some visible B center activations (b) CL map in the vicinity of a preexisting B center (c) Radial intensity of the data in (a) (blue dots) fitted to a Gaussian profile (green curve).}%
 \end{figure}

In this section, we provide information regarding the shape of the CL spots of B center emission in the scanning regime. Regarding the apparent size of the B center in CL scans, we have obtained an average FWHM value of $262 \pm 47$~nm for a total 29 centers. Figure \ref{CL shape}a displays a CL map with multiple B center activations, as well as B centers already present and one B center deactivation as observed in our previous work~\cite{Nunez26}, while Figure \ref{CL shape}b shows a CL map of a single B center that is already activated. To verify that the CL intensity distribution function is Gaussian, we integrate the signal radially with respect to the center of the spot. The blue dots on Fig~\ref{CL shape}c show the resulting signal as a function of the radius $r$. The line shape is well fitted to a Gaussian function (green curve).

\section*{S5. Estimation of the color center density}
\begin{figure}[h]
 \includegraphics[width=\linewidth]{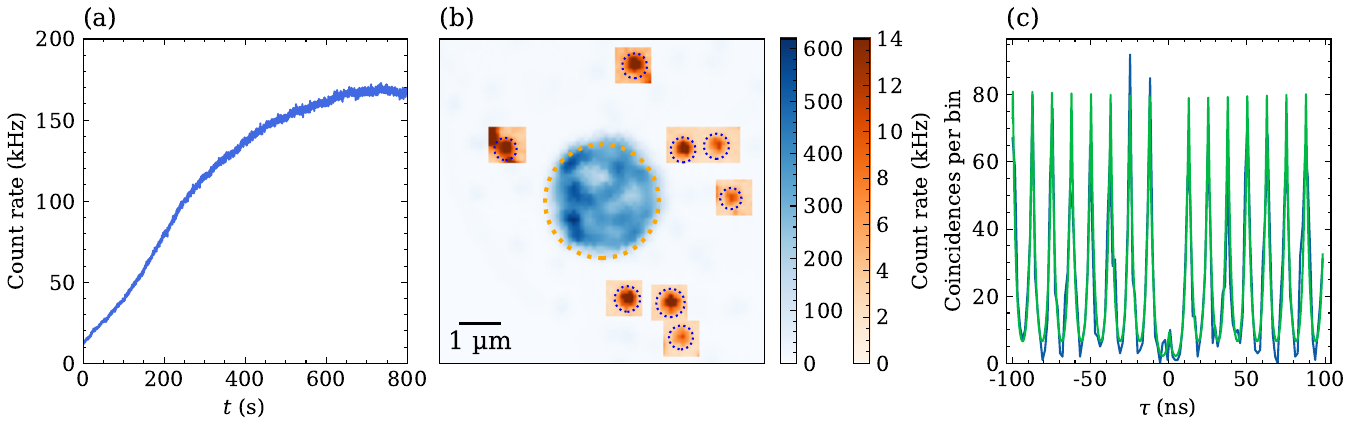}%
 \caption{\label{Density} (a) CL time trace recorded during electron irradiation on the hBN sample, exhibiting saturation of SPE activation. (b) PL map of the irradiated area with an ensemble of B centers (blue scale). A secondary color scale (red) reveals individual B centers. (c) Blue line: $g^{(2)}(\tau)$ measured on one of the individual B centers represented with the orange colorbar in (b). Green line: fit to the data, providing $g^{(2)}(0) = 0.11 \pm 0.10$.}%
 \end{figure}

In this section, we estimate the maximum density of color centers that can be created per unit volume, which we associate with the initial density of UV-emitting color centers. The approach follows Ref.~\cite{Roux22}. We first perform a continuous irradiation of the sample with a defocused beam of approximately 1~$\mathrm{\mu m}$ radius. The irradiation parameters are $V = 3.0$~kV and $I = 0.8$~nA. We keep the beam on until the CL emission, shown in Figure \ref{Density}a, reaches a plateau, indicating the saturation of B center creation. 

We then perform PL confocal mapping of the irradiated area. As shown on Fig.~\ref{Density}b, we observe a central region with a high density of emitters (blue scale), where the beam directly irradiated the sample, accompanied by single emitters in the vicinity (red scale). The latter are verified to be single emitters through measurement of the second-order intensity correlation function $g^{(2)}(\tau)$ (Fig.~\ref{Density}c). The estimated number of emitters in the central area $N$ is given by the ratio between the total integrated intensity of the ensemble and the integrated intensity of individual B centers. Then, knowing the volume of the irradiated zone $V_\mathrm{spot}$ -- which is given spot area multiplied by the flake thickness --, we calculate the density as $n_{v} = N/ V_{spot}$. Repeating this process on 5 spots, we obtain a value of $3.6 \pm 0.6 \times 10^{15}$~cm$^{-3}$.

\begin{thebibliography}{99}
\bibitem{Aharonovich2016} Aharonovich, I.; Englund, D.; Toth, M. Solid-State Single-Photon Emitters. \textit{Nat. Photonics} 2016, \textbf{10}, 631--641. https://doi.org/10.1038/nphoton.2016.186

\bibitem{Zhang20} Zhang, G.; Cheng, Y.; Chou, J.-P.; Gali, A. Material Platforms for Defect Qubits and Single-Photon Emitters. \textit{Appl. Phys. Rev.} 2020, \textbf{7}, 031308. https://doi.org/10.1063/5.0006075

\bibitem{Tran16} Tran, T. T.; Bray, K.; Ford, M. J.; Toth, M.; Aharonovich, I. Quantum Emission from Hexagonal Boron Nitride Monolayers. \textit{Nat. Nanotechnol.} 2016, \textbf{11}, 37--41. https://doi.org/10.1038/nnano.2015.242

\bibitem{Martinez16} Mart\'inez, L. J.; Pelini, T.; Waselowski, V.; Maze, J. R.; Gil, B.; Cassabois, G.; Jacques, V. Efficient Single Photon Emission from a High-Purity Hexagonal Boron Nitride Crystal. \textit{Phys. Rev. B} 2016, \textbf{94}, 121405. https://doi.org/10.1103/PhysRevB.94.121405

\bibitem{Vogl25} Çakan, A.; Cholsuk, C.; Gale, A.; Kianinia, M.; Paçal, S.; Ateş, S.; Aharonovich, I.; Toth, M.; Vogl, T. Quantum Optics Applications of Hexagonal Boron Nitride Defects. \textit{Adv. Opt. Mater.} 2025, \textbf{13}, 2402508. https://doi.org/10.1002/adom.202402508

\bibitem{Shevitski19} Shevitski, B.; Gilbert, S. M.; Chen, C. T.; Kastl, C.; Barnard, E. S.; Wong, E.; Ogletree, D. F.; Watanabe, K.; Taniguchi, T.; Zettl, A.; Aloni, S. Blue-Light-Emitting Color Centers in High-Quality Hexagonal Boron Nitride. \textit{Phys. Rev. B} 2019, \textbf{100}, 155419. https://doi.org/10.1103/PhysRevB.100.155419

\bibitem{Fournier21} Fournier, C.; Plaud, A.; Roux, S.; Pierret, A.; Rosticher, M.; Watanabe, K.; Taniguchi, T.; Buil, S.; Qu\'elin, X.; Barjon, J.; Hermier, J.-P.; Delteil, A. Position-Controlled Quantum Emitters with Reproducible Emission Wavelength in Hexagonal Boron Nitride. \textit{Nat. Commun.} 2021, \textbf{12}, 3779. https://doi.org/10.1038/s41467-021-24019-6

\bibitem{Gale22} Gale, A.; Li, C.; Chen, Y.; Watanabe, K.; Taniguchi, T.; Aharonovich, I.; Toth, M. Site-Specific Fabrication of Blue Quantum Emitters in Hexagonal Boron Nitride. \textit{ACS Photonics} 2022, \textbf{9}, 2170--2177. https://doi.org/10.1021/acsphotonics.2c00631

\bibitem{Nunez26} Núñez Marcos, A.; Arnold, C.; Barjon, J.; Buil, S.; Hermier, J.-P.; Delteil, A. Deterministic Generation of Single B Centers in hBN by One-to-One Conversion from UV Centers. \textit{ACS Nano} 2026, \textbf{20} (13), 10404--10411. https://doi.org/10.1021/acsnano.5c19102

\bibitem{Horder22} Horder, J.; White, S. J. U.; Gale, A.; Li, C.; Watanabe, K.; Taniguchi, T.; Kianinia, M.; Aharonovich, I.; Toth, M. Coherence Properties of Electron-Beam-Activated Emitters in Hexagonal Boron Nitride under Resonant Excitation. \textit{Phys. Rev. Appl.} 2022, \textbf{18}, 064021. https://doi.org/10.1103/PhysRevApplied.18.064021

\bibitem{Fournier23PRA} Fournier, C.; Roux, S.; Watanabe, K.; Taniguchi, T.; Buil, S.; Barjon, J.; Hermier, J.-P.; Delteil, A. Two-Photon Interference from a Quantum Emitter in Hexagonal Boron Nitride. \textit{Phys. Rev. Appl.} 2023, \textbf{19}, L041003. https://doi.org/10.1103/PhysRevApplied.19.L041003

\bibitem{Gerard24} G\'erard, D.; Pierret, A.; Fartas, H.; Berini, B.; Buil, S.; Hermier, J.-P.; Delteil, A. Quantum Efficiency and Vertical Position of Quantum Emitters in hBN Determined by Purcell Effect in Hybrid Metal-Dielectric Planar Photonic Structures. \textit{ACS Photonics} 2024, \textbf{11}, 5188--5194. https://doi.org/10.1021/acsphotonics.4c01416

\bibitem{Gerard26} G\'erard, D.; Buil, S.; Watanabe, K.; Taniguchi, T.; Hermier, J.-P.; Delteil, A. Resonance Fluorescence and Indistinguishable Photons from a Coherently Driven B Centre in hBN. \textit{Nat. Commun.} 2026, \textbf{17}, 1843. https://doi.org/10.1038/s41467-026-68555-5

\bibitem{Gerard23} G\'erard, D.; Rosticher, M.; Watanabe, K.; Taniguchi, T.; Barjon, J.; Buil, S.; Hermier, J.-P.; Delteil, A. Top-Down Integration of a hBN Quantum Emitter in a Monolithic Photonic Waveguide. \textit{Appl. Phys. Lett.} 2023, \textbf{122}, 264001. https://doi.org/10.1063/5.0152721

\bibitem{Spencer23} Spencer, L.; Horder, J.; Kim, S.; Toth, M.; Aharonovich, I. Monolithic Integration of Single Quantum Emitters in hBN Bullseye Cavities. \textit{ACS Photonics} 2023, \textbf{10}, 4417--4424. https://doi.org/10.1021/acsphotonics.3c01282

\bibitem{Mackoit19} Mackoit-Sinkevi\v{c}ien\.{e}, M.; Maciaszek, M.; Van de Walle, C. G.; Alkauskas, A. Carbon Dimer Defect as a Source of the 4.1 eV Luminescence in Hexagonal Boron Nitride. \textit{Appl. Phys. Lett.} 2019, \textbf{115}, 212101. https://doi.org/10.1063/1.5124153

\bibitem{Li22} Li, S.; Pershin, A.; Thiering, G.; Udvarhelyi, P.; Gali, A. Ultraviolet Quantum Emitters in Hexagonal Boron Nitride from Carbon Clusters. \textit{J. Phys. Chem. Lett.} 2022, \textbf{13}, 3150--3157. https://doi.org/10.1021/acs.jpclett.2c00665

\bibitem{Plo25} Plo, J.; Pershin, A.; Li, S.; Poirier, T.; Janzen, E.; Schutte, H.; Tian, M.; Wynn, M.; Bernard, S.; Rousseau, A.; Ibanez, A.; Valvin, P.; Desrat, W.; Michel, T.; Jacques, V.; Gil, B.; Kaminska, A.; Wan, N.; Edgar, J. H.; Gali, A.; Cassabois, G. Isotope Substitution and Polytype Control for Point Defects Identification: The Case of the Ultraviolet Color Center in Hexagonal Boron Nitride. \textit{Phys. Rev. X} 2025, \textbf{15}, 021045. https://doi.org/10.1103/PhysRevX.15.021045

\bibitem{Roux22} Roux, S.; Fournier, C.; Watanabe, K.; Taniguchi, T.; Hermier, J.-P.; Barjon, J.; Delteil, A. Cathodoluminescence Monitoring of Quantum Emitter Activation in Hexagonal Boron Nitride. \textit{Appl. Phys. Lett.} 2022, \textbf{121}, 184002. https://doi.org/10.1063/5.0126357

\bibitem{Nedic24} Nedi\'c, S.; Yamamura, K.; Gale, A.; Aharonovich, I.; Toth, M. Electron Beam Restructuring of Quantum Emitters in Hexagonal Boron Nitride. \textit{Adv. Opt. Mater.} 2024, \textbf{12}, 2400908. https://doi.org/10.1002/adom.202400908

\bibitem{Hou25} Hou, H.; Hua, M.; Kolluru, V. S. C.; Chen, W.-Y.; Yin, K.; Tripathi, P.; Chan, M. K. Y.; Diroll, B. T.; Gage, T. E.; Zuo, J.-M.; Wen, J. Nanometer Resolution Structure-Emission Correlation of Individual Quantum Emitters via Enhanced Cathodoluminescence in Twisted Hexagonal Boron Nitride. \textit{Adv. Mater.} 2025, \textbf{37}, e01611. https://doi.org/10.1002/adma.202501611

\bibitem{Liu25} Liu, L.; Khanonkin, I.; Eberle, J.; Rizek, B.; F\"alt, S.; Watanabe, K.; Taniguchi, T.; Imamo\u{g}lu, A.; Kroner, M. Single Photon Emitters in Ultra-Thin Hexagonal Boron Nitride Layers. \textit{arXiv} 2025, arXiv:2507.02633. https://doi.org/10.48550/arXiv.2507.02633

\bibitem{Taniguchi07} Taniguchi, T.; Watanabe, K. Synthesis of High-Purity Boron Nitride Single Crystals Under High Pressure by Using Ba--BN Solvent. \textit{J. Cryst. Growth} 2007, \textbf{303}, 525--529. https://doi.org/10.1016/j.jcrysgro.2006.12.061

\bibitem{Laurent2026} Laurent, E.; Pe\~na Rom\'an, R. J.; Miller, S.; Moghe, A. R.; Lorchat, E.; Le Moal, S.; Boer-Duchemin, E.; Zagonel, L. F.; Berciaud, S.; Le Moal, E. Sub-Diffraction-Resolved Spatial Distribution of Emitting Excitons in Scanning Tunneling Microscopy-Induced Luminescence of Two-Dimensional Semiconductors via Richardson--Lucy Deconvolution. \textit{Phys. Rev. Appl.} 2026, \textbf{26}, 014016. https://doi.org/10.1103/lwl7-zrk8

\bibitem{LamraniAlaoui2026} Lamrani Alaoui, A.; Moreno, \'A.; Heithoff, M.; Br\"andli, V.; Courville, A.; Gromovyi, M.; Chenot, S.; Srivastava, M. S.; V\'ezian, S.; Damilano, B.; Koppens, F.; Chassagneux, Y.; Voisin, C.; Boucaud, P.; Reserbat-Plantey, A. Strain and Interface-Driven Localized Emission in MoSe$_2$ Monolayers on GaN Nanopillars. \textit{ACS Photonics} 2026, \textbf{13}, 5693--5702. https://doi.org/10.1021/acsphotonics.6c01384

\bibitem{Liu17} Liu, J.; Davanço, M. I.; Sapienza, L.; Konthasinghe, K.; De Miranda Cardoso, J. V.; Song, J. D.; Badolato, A.; Srinivasan, K. Cryogenic Photoluminescence Imaging System for Nanoscale Positioning of Single Quantum Emitters. \textit{Rev. Sci. Instrum.} 2017, \textbf{88}, 023116. https://doi.org/10.1063/1.4976578

\bibitem{Sartison17} Sartison, M.; Portalupi, S. L.; Gissibl, T.; Jetter, M.; Giessen, H.; Michler, P. Combining In Situ Lithography with 3D Printed Solid Immersion Lenses for Single Quantum Dot Spectroscopy. \textit{Sci. Rep.} 2017, \textbf{7}, 39916. https://doi.org/10.1038/srep39916

\bibitem{Chen19} Chen, Y.-C.; Griffiths, B.; Weng, L.; Nicley, S. S.; Ishmael, S. N.; Lekhai, Y.; Johnson, S.; Stephen, C. J.; Green, B. L.; Morley, G. W.; Newton, M. E.; Booth, M. J.; Salter, P. S.; Smith, J. M. Laser Writing of Individual Nitrogen-Vacancy Defects in Diamond with Near-Unity Yield. \textit{Optica} 2019, \textbf{6}, 662--667. https://doi.org/10.1364/OPTICA.6.000662

\bibitem{Chen17} Chen, Y.-C.; Salter, P. S.; Knauer, S.; Weng, L.; Frangeskou, A. C.; Stephen, C. J.; Ishmael, S. N.; Dolan, P. R.; Johnson, S.; Green, B. L.; Morley, G. W.; Newton, M. E.; Rarity, J. G.; Booth, M. J.; Smith, J. M. Laser Writing of Coherent Colour Centres in Diamond. \textit{Nat. Photonics} 2017, \textbf{11}, 77--80. https://doi.org/10.1038/nphoton.2016.234

\bibitem{Kim25} Kim, S.; London, P.; Yang, D.; Hughes, L. B.; Ahlers, J.; Meynell, S.; Mitchell, W. J.; Mukherjee, K.; Bleszynski Jayich, A. C. Scalable Nanoscale Positioning of Highly Coherent Color Centers in Prefabricated Diamond Nanostructures. \textit{Nat. Commun.} 2025, \textbf{16}, 9803. https://doi.org/10.1038/s41467-025-64758-4

\bibitem{Cheng25} Cheng, X.; Thurn, A.; Chen, G.; Jones, G. S.; Bennett, J. E.; Coke, M.; Adshead, M.; Michaels, C. P.; Balci, O.; Ferrari, A. C.; Atatüre, M.; Curry, R. J.; Smith, J. M.; Salter, P. S.; Gangloff, D. A. Laser Activation of Single Group-IV Colour Centres in Diamond. \textit{Nat. Commun.} 2025, \textbf{16}, 5124. https://doi.org/10.1038/s41467-025-60373-5

\bibitem{Wang17} Wang, J.; Zhang, X.; Zhou, Y.; Li, K.; Wang, Z.; Peddibhotla, P.; Liu, F.; Bauerdick, S.; Rudzinski, A.; Liu, Z.; Gao, W. Scalable Fabrication of Single Silicon Vacancy Defect Arrays in Silicon Carbide Using Focused Ion Beam. \textit{ACS Photonics} 2017, \textbf{4}, 1054--1059. https://doi.org/10.1021/acsphotonics.7b00230

\bibitem{He23} He, Z.-X.; Li, Q.; Wen, X.-L.; Zhou, J.-Y.; Lin, W.-X.; Hao, Z.-H.; Xu, J.-S.; Li, C.-F.; Guo, G.-C. Maskless Generation of Single Silicon Vacancy Arrays in Silicon Carbide by a Focused He$^{+}$ Ion Beam. \textit{ACS Photonics} 2023, \textbf{10}, 2234--2240. https://doi.org/10.1021/acsphotonics.2c01209

\bibitem{Hollenbach22} Hollenbach, M.; Klingner, N.; Jagtap, N. S.; Bischoff, L.; Fowley, C.; Kentsch, U.; Hlawacek, G.; Erbe, A.; Abrosimov, N. V.; Helm, M.; Berencén, Y.; Astakhov, G. V. Wafer-Scale Nanofabrication of Telecom Single-Photon Emitters in Silicon. \textit{Nat. Commun.} 2022, \textbf{13}, 7683. https://doi.org/10.1038/s41467-022-35051-5

\bibitem{Groot19} Groot-Berning, K.; Kornher, T.; Jacob, G.; Stopp, F.; Dawkins, S. T.; Kolesov, R.; Wrachtrup, J.; Singer, K.; Schmidt-Kaler, F. Deterministic Single-Ion Implantation of Rare-Earth Ions for Nanometer-Resolution Color-Center Generation. \textit{Phys. Rev. Lett.} 2019, \textbf{123}, 106802. https://doi.org/10.1103/PhysRevLett.123.106802

\bibitem{CASINO} Demers, H.; Poirier-Demers, N.; Réal Couture, A.; Joly, D.; Guilmain, M.; de Jonge, N.; Drouin, D. Three-Dimensional Electron Microscopy Simulation with the CASINO Monte Carlo Software. \textit{Scanning} 2011, \textbf{33} (3), 135--146. https://doi.org/10.1002/sca.20262

\bibitem{Pelini19} Pelini, T.; Elias, C.; Page, R.; Xue, L.; Liu, S.; Li, J.; Edgar, J. H.; Dréau, A.; Jacques, V.; Valvin, P.; Gil, B.; Cassabois, G. Shallow and Deep Levels in Carbon-Doped Hexagonal Boron Nitride Crystals. \textit{Phys. Rev. Mater.} 2019, \textbf{3}, 094001. https://doi.org/10.1103/PhysRevMaterials.3.094001

\bibitem{Williams25} Williams, E.; Gale, A.; Horder, J.; Scognamiglio, D.; Toth, M.; Aharonovich, I. Quantum Emitters in Flux Grown hBN. \textit{Cryst. Growth Des.} 2025, \textbf{25}, 2083--2089. https://doi.org/10.1021/acs.cgd.4c01713

\bibitem{Alemoush26} Alemoush, Z.; Almohammad, M.; Li, J.; Lin, J. Y.; Jiang, H. X. Carbon-Doped Quasi-Bulk Hexagonal Boron Nitride Grown by HVPE. \textit{Appl. Phys. Lett.} 2026, \textbf{128}, 152106. https://doi.org/10.1063/5.0326417

\bibitem{SIRAF} Anders Brostrøm, Kristian Mølhave, Spatial Image Resolution Assessment by Fourier Analysis (SIRAF), Microscopy and Microanalysis, Volume 28, Issue 2, 1 April 2022, Pages 469–477 


\bibitem{hBNdensity} Madelung, O., Semiconductors: Group IV elements and III-V compounds. Series "Data in science and technology", Springer Berlin Heidelberg, 1991, pag. 164

\end{thebibliography}

\begin{thebibliography}{99}







\end{thebibliography}
\end{document}